# Focusing of transition radiation and diffraction radiation from concave targets.

A.P. Potylitsyn[*], R.O. Rezaev

*Tomsk Polytechnic University, pr. Lenina 30, 634050, Tomsk, Russia*

**Abstract**

In the article the transition radiation from a relativistic charge crossing a finite radius paraboloid target with taking into account the pre-wave zone effect is considered. It is shown that in this case the radiation cone narrowing occurs in contrast with the transition radiation from the flat target if the detector is situated at the distance, which is smaller than the focus distance (the focusing effect) from the target. At the charged particle passage through the central hole in a paraboloid target the diffraction radiation focusing (DR) occurs too. The focusing of coherent DR for the non-invasive measuring of the electron bunch length is proposed.

1. Transition radiation (TR) is widely used for an electron beam diagnostics. For instance, in the experiment [1] for measurements of the transverse size of an electron beam, the optical transition radiation focusing system allowed the spatial resolution to be reached at about ~ 5 mcm. In the work [2] an electron bunch length was measured using the spectrum of coherent transition radiation (CTR). The alternative approach was demonstrated by the authors of work [3], where the CTR beam was being generated by the bunch of the length of about $l_b \sim$ 1mm, then it was being focused by the complicated optical system on the electro optical crystal, through which the linear-polarized radiation of a stable laser passed. The crystal becomes birefringence under the influence of electric field $\vec{E}_{CTR}$ that leads to the circularly-polarized component occurrence in the laser radiation after the crystal with degree $\Delta P_C$ the value of

---
[*]Corresponding authors:
*E-mail address*: pap@interact.phtd.tpu.edu.ru

which is in proportion to the field $\vec{E}_{CTR}$ and spill time $\Delta t$:

$$\Delta P_C \sim \left|\vec{E}_{CTR}\right|\Delta t \sim \left|\vec{E}_{CTR}\right| l_b/c \qquad (1)$$

The bunch length $l_b$ was determined from the value $\Delta P_C$ measured. The using of TR – target for diagnostics leads to a beam emittance growth in avoidable.

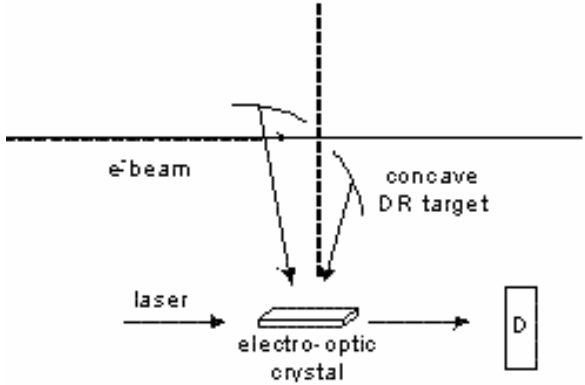

Fig. 1. Scheme of experiment

The non-invasive diagnostics based on the diffraction radiation use was developed in the works [4], [5]. We can suggest the diagnostics scheme, which is analogous to the scheme used in the experiment [3] where instead TR, the diffraction radiation (DR) is generated by the electron beam passing through the hole of the concave target (see Fig. 1). In this case, firstly, the diagnostics becomes practically non-invasive, secondly, DR focusing by the target itself simplifies the experimental equipment (there is no necessity to have a focusing system) and thirdly, increases the technique sensitivity because the field intensity of focused radiation from the concave target exceeds the field intensity from the flat one.

Actually, as it was mentioned in work [6], the virtual photon field of ultrarelativistic charge is close to transverse field of electromagnetic wave by its characteristics. The virtual photon field, in analogy with electromagnetic wave, bends around the screen of the finite sizes, that is the origin of DR occurrence. So, we can suppose that for the parabolic target the focusing effect in transition and diffraction radiation of ultrarelativistic particles must exist.

In the series of works [7], [8], [9], the models for TR characteristics calculation were developed for the deformed targets of the finite sizes. In our paper, the approach which allows taking into account the influence of the finite size [10], [11], the pre-wave zone effect [12] and the "global" form of the deformed target (for examples, paraboloid), but not a "local" one as in works [7], [9] has been developed.

Besides, the suggested approach allows both the characteristic of TR and DR to be calculated. As in the cited works, the perfectly conducting target is considered.

2. Let us consider the diffraction radiation in the pre-wave zone using the model developed in [13]. For the simplicity of calculations, let's choose the geometry when the relativistic electron with the constant velocity $v$ passes near the semi-infinite plane with almost zero impact parameter when the target is inclined from the perpendicular position at angle of $\alpha \sim \gamma^{-1} \ll 1$ (see Fig. 2). The field of the backward DR may be calculated on the detector plane from the expression:

$$\begin{Bmatrix} E_x^D(X_D,Y_D) \\ E_y^D(X_D,Y_D) \end{Bmatrix} = const \int dX_T dY_T \begin{Bmatrix} X_T \\ Y_T \end{Bmatrix} \frac{K_1\left(\frac{k}{\beta\gamma}\sqrt{X_T^2+Y_T^2}\right)}{\sqrt{X_T^2+Y_T^2}} \times \exp[i\Delta\varphi], \quad (2)$$

In (2) variables with indexes $T, D$ are described the coordinates reference in the system on the target and detector surfaces respectively, $\Delta\varphi$ - is the radiation phase shift, $K_1$ – is the second kind Bessel modified function, the $k = \frac{2\pi}{\lambda}$ - is the wave vector, $\lambda$ - is the radiation wave length, $\beta = v/c$, $\gamma$ - is the Lorentz-factor.

The calculations will be made in the dimensionless variables:

$$\begin{Bmatrix} x_T \\ y_T \end{Bmatrix} = \frac{2\pi}{\gamma\lambda} \begin{Bmatrix} X_T \\ Y_T \end{Bmatrix}, \quad \begin{Bmatrix} x_D \\ y_D \end{Bmatrix} = \frac{\gamma}{L} \begin{Bmatrix} X_D \\ Y_D \end{Bmatrix}, \quad R = \frac{L}{\gamma^2\lambda}, \quad (3)$$

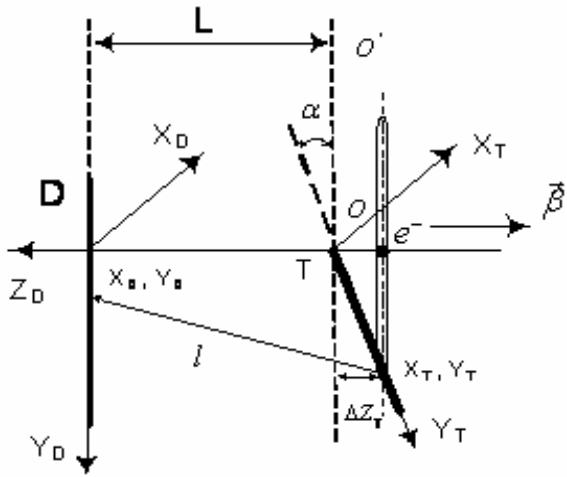

Fig. 2. To the calculation of phase shift; the electron passes near target with zero impact parameter.

here $L$ – is the distance from the target to the detector.

The integration in (2) is performed on the target surface, which can be limited.

As it has been mentioned in article [12], the "DR source" at the distance $L \leq \gamma^2\lambda$, can't be considered on the target surface as a point-wise one. In other words, for large-size dimensionless parameter

$R = L/\gamma^2\lambda \gg 1$, the DR (or TR) characteristics can be described as angular distributions from the point-wise source (so-called far zone), while at $R \leq 1$ the "pre-wave zone effect" [12] plays the important role in the cases where the natural size of the luminous" target area, which is defined order by the size parameter $\gamma\lambda$, should be taken into account. In this case (especially at $R \ll 1$) it is more natural to describe DR (TR) characteristics as coordinate distributions, for example, on the detector's surface.

The phase shift $\Delta\varphi$ in the expression (2) will be calculated relative to the plane $OO'$ (see Fig. 2).

$$\Delta\varphi = \frac{2\pi\left(l+\frac{\Delta Z_T}{\beta}\right)}{\lambda} = \frac{2\pi}{\lambda}\sqrt{(Y_T\cos\alpha - Y_D)^2 + (X_T - X_D)^2 + (L+\Delta Z_T(X_T,Y_T))^2} + \\ + \frac{2\pi}{\lambda}\frac{\Delta Z_T(X_T,Y_T)}{\beta}, \quad (4)$$

here $l$ - is the distance between the point on the target $(X_T,Y_T)$ (the point where the spherical wave emitted) and the point on the detector $(X_D,Y_D)$, $\Delta Z_T = Y_T\sin\alpha$ (see Fig. 2). The summand with $\Delta Z_T$ in the expression (4), takes into account the fact that the transverse field of particle (virtual photon source) requires the finite time, which is counted from plane $OO'$, for achieving the target point $(X_T,Y_T)$, after that the virtual photons are transformed into the real ones and spherical wave is emitted. After the expansion of the expression (3) in the series by small parameters $\left(\frac{X_T}{L},\frac{Y_T}{L},\frac{Y_D}{L},\frac{Y_D}{L}\right)$ and the dimensionless variables (2) usage, we'll get the following phase shift (neglecting the terms having the constant parameters):

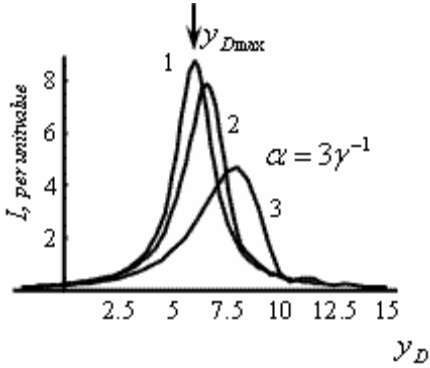

Fig. 3. DR intensity distribution on the detector surface from an inclined flat target for different values of parameter $R$ (zero impact parameter):
$1 - R = 10; 2 - R = 0.5; 3 - R = 0.1$.

$$\exp[i\Delta\varphi] = \exp\left[i\left(\frac{x_T^2 + y_T^2\cos^2\alpha}{4\pi R}\right)\right] \times \exp\left[i\pi R(x_D^2 + y_D^2)\right] \times \\ \times \exp\left[-i(y_T y_D\cos\alpha + x_T x_D)\right] \times \exp\left[i\gamma\, y_T\sin\alpha\left(1+\frac{1}{\beta}\right)\right]. \quad (5)$$

In Figure 3 the distribution of diffraction radiation intensity from the inclined target $(\alpha = 3\gamma^{-1})$ on the detector plane (at $X_D = 0$) at the zero impact parameter, for the distance $L$ between it and the detector (see Fig. 2) is shown. The radiation intensity accurate to a constant is determined by the expression: $I \sim |E_x^D|^2 + |E_y^D|^2$. It is evident from figure 3, that in the far zone the symmetrical distribution of intensity (curve 1) is observed, the maximum position being

corresponded to the value $y_{D\max} = 6$, that is, in other words, the distribution maximum coincides with the direction of the specular reflection.

It should be noticed that the obtained distribution coincides with the well-known angular distribution DR [11]:

$$\frac{dI}{d\Omega} = const \frac{1 + x_D^2}{\left(1 + y_D^2\right)\left(1 + x_D^2 + y_D^2\right)}, \qquad (6)$$

where variables $x_D, y_D$ in the far zone correspond to the angular variable $\gamma\theta_x, \gamma\theta_y$, where the angles $\theta_x, \theta_y$ are counted from the specular reflection direction.

In the pre-wave zone $(R < 1)$, the asymmetric distribution (curves 2, 3) is observed, the distribution maximum being shifted to the side of large angles ($y_{D\max} > 2\gamma\alpha$). At the DR generated in the target with the variable inclination angle (for example, in paraboloid), if the detector is situated in the pre-wave zone, the similar behavior of the "reflected" photons groups them near the paraboloid axis, that is, the radiation focusing occurs (see below).

3. To begin, let us consider the focusing effect on the example of transition radiation for the perfectly reflecting target made in the form of paraboloid, along the axis of which the charged particle moves (to the positive direction of axis $Z_T$, see Fig. 4). Let's consider the geometry of backwards transition radiation again, as it was mentioned above. The surface of such target is described by the equation:

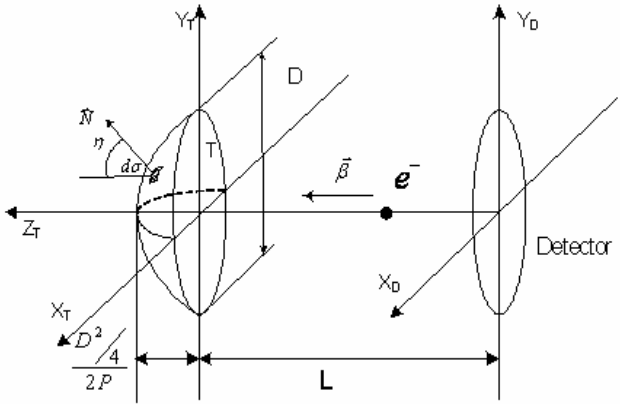

Fig. 4. The geometry of the focusing target.
$\vec{N}$ - vector of normal to the surface;
$d\sigma$ is an elementary area.

$$2PZ_T = D^2/4 - Y_T^2 - X_T^2, \qquad (7)$$

$D$ – is the maximum target diameter,
$P$ - is the focusing distance, $L$ – is the distance from the target borders to the detector (Fig. 4).

We'll consider the "radial" component of TR field $E_r^D$ because of azimuthally symmetry. For the flat target of diameter $D$, the radial component is calculated by the following manner [13]:

$$E_r^D(\vec{R}_D) = const \int_0^{d/2} r_T dr_T K_1(r_T) \times$$
$$\times \exp\left[i\gamma\left(\frac{d^2/4 - r_T^2}{2p}\right)\left(1 + \frac{1}{\beta}\right)\right] \exp\left[i\frac{r_T^2}{4\pi R}\right] \exp[i\pi R r_D^2] J_1(r_T r_D). \quad (8)$$

Here, as it was mentioned above, the dimensionless variables are used:

$$r_T = \frac{2\pi R_T}{\gamma\lambda}, \quad r_D = \frac{\gamma}{L} R_D, \quad d = \frac{2\pi D}{\gamma\lambda}, \quad p = \frac{2\pi P}{\gamma\lambda}, \quad (9)$$

where $R_{T,D} = \sqrt{X_{T,D}^2 + Y_{T,D}^2}$ ( the capital letter stands the dimensional variable, the small letter stands the dimensionless one).

At the analogous calculation of field $E_r^D$ from the concave target the integration is performed by the paraboloid surface, the area element of which is given by the formula:

$$d\sigma = \frac{dX_T dY_T}{|\cos\eta|} = dX_T dY_T \frac{\sqrt{X_T^2 + Y_T^2 + P^2}}{P} = \frac{2\pi\sqrt{R_T^2 + P^2}}{P} R_T dR_T, \quad (10)$$

where $\eta$ - is the angle between the normal $\vec{N}$ of the surface in point $(X_T, Y_T)$ and the axis $Z_T$ (see Fig. 4). The equation (10) must be used instead of $r_T dr_T$ in the dimensionless variables in the integral (8).

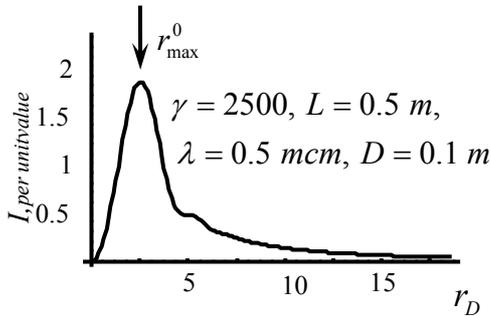

Fig. 5 a. The radial distribution of TR intensity from the flat circular target $R = 0.16$.

Let's compare the TR intensity from flat and concave targets on the detector placed in pre-wave zone if the diameter of parabolic target is $D \gg \gamma\lambda$.

In Fig. 5 a the distribution of TR intensity on the detector from a target for $R = 0.16$ (that is, $L=0.16\gamma^2\lambda$) and $d=500$ (or in dimension units $D = \frac{\gamma\lambda}{2\pi}d \approx 80\gamma\lambda$) is shown and in Fig. 5 b the evolution of TR distributions from parabolic target with changing focus( $P=2L$ – curve 1, $P=L$ – curve 2, $P=0.33L$ – curve 3) are shown

As it follows from the obtained results, if the detector is situated at the distance $L < P$ before the mirror focus, the radiation focusing is observed, for the focus position the distributions coincide from flat and parabolic targets and the defocusing is observed if $L>P$.

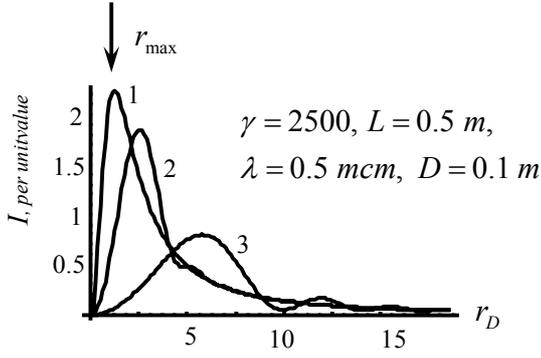

Fig. 5 b. The radial distribution of TR intensity from paraboloid target for different values of focusing distance (curve 1 - $p=5000$; curve 2 - $p=2500$; curve 3 - $p=800$); $R = 0.16$.

Using the term "focusing" we shall understand the cone narrowing of the radiation from paraboloid target in comparison with the radiation from flat one, that is, $r_{max} < r_{max}^0$ (the position of maximum in the radial distribution of TR intensity from concave and flat target is symbolized by $r_{max}, r_{max}^0$).

It may be shown that the energy of TR generated in the target by relativistic particle doesn't depend on whether this target is flat or deformed, and is determined by the Lorentz factor of the particle and by the target diameter.

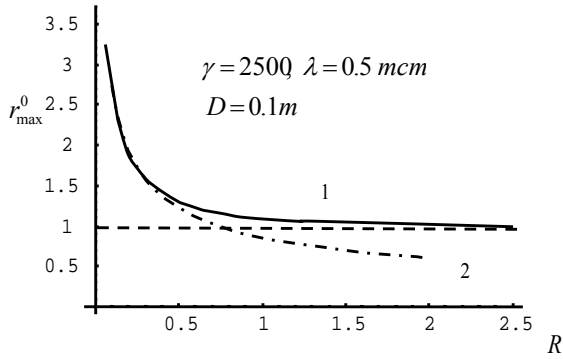

Fig. 6 a. The dependence on maximum position of TR intensity on the detector from a flat target (solid line); approximation (11) – dashed-dotted line; approximation (12) – dashed line.

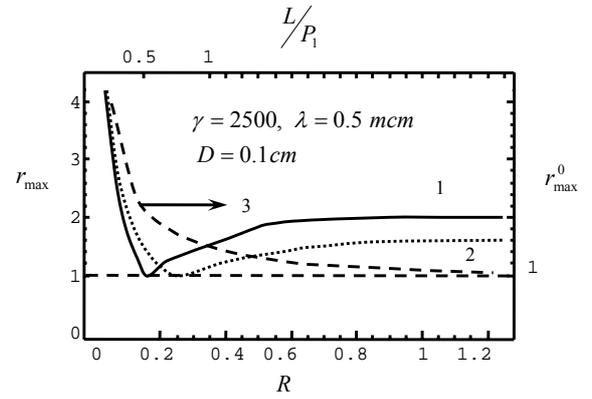

Fig. 6 b. The dependence on the maximum position of TR intensity from paraboloid target for various focus distance: $P_1 = 0.32\gamma^2\lambda$ (curve 1 – solid line), $P_2 = 0.5\gamma^2\lambda$ (curve 2 – dotted line); the same dependence for the flat target – curve 3 (dashed line) is given for the comparison.

The dependence of maximum position on the distance between the flat target and the detector $r_{max}^0(R)$, where the distance is taken in units $\gamma^2\lambda$, is shown in Fig. 6 a. For the values $R << 1 \; (L << \gamma^2\lambda)$ the position of maximum is approximated by the function [13]:

$$r_{max}^0(R) \approx \frac{0.863}{\sqrt{R}}, \qquad (11)$$

while the asymptotic behavior at $R > 1$ is the following:

$$r_{max}^0 (R) = 1, \qquad (12)$$

as expected because in the far zone the maximum in TR distribution corresponds to the angle $\theta_{max} = \gamma^{-1}$, that is, $R_{max}^0 = a/\gamma$ $(r_{max}^0 = 1)$. Approximation (11) agrees well with the result of exact calculations for the values $R \leq 0.3$. In Fig. 6 b the analogous dependences for the paraboloid target with $P=0.33\,\gamma^2\lambda$ (curve 1) and $P=\gamma^2\lambda$ (curve 2) are shown. It can be noticed that the maximal focusing effect is achieved at the distance $L = P/2$, while at $L > P$ the defocusing is observed. Thus, the minimum value of the quantity $r_{max}$ in the point $L = P/2$ achieves the unity. In analogy with optics it can be expected that the conclusion is right for the targets of large diameter $(D \gg \gamma\lambda)$.

In Fig. 7, the dependence of radius $r_{max}$ on the paraboloid diameter is shown. The given behavior is universal because it doesn't depend on the parameter $R$ (the calculation for $R$ in the range from 0,001 to 1 coincides with the precision, which is better than $10^{-3}$ %).

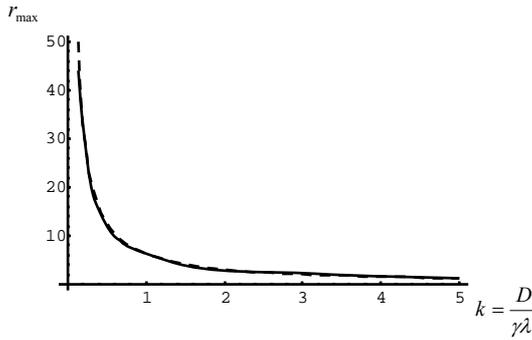

Fig. 7. The dependence on radius of the focused spot on the detector on the sizes of target diameter $D$ in TR distribution (solid curve); the approximation (14) is given by the dashed line.

Making the further analogy with optics, it can be supposed that the "spot" size of the focusing radiation $R_{opt}$ is determined by the wave length $\lambda$ and aperture of the focusing mirror $\theta_{ap} = \dfrac{D}{2L}$:

$$R_{opt} \sim \frac{\lambda}{\theta_{ap}} = \frac{2\pi L\lambda}{D}.$$

If in our case the diameter of target may be given in the form:

$$D = k\gamma\lambda,\ k \leq 1,$$

it can be expected that the spot size of the focused TR is described by formula

$$R_{max} \approx 2\pi \frac{L}{\gamma} \frac{1}{k}, \qquad (13)$$

or considering the dimensionless coordinates, we can give the following formula:

$$r_{max} \approx \frac{2\pi}{k}. \qquad (14)$$

In Figure 7, the last dependence is shown by the dashed line.

4. Let's consider the focusing of transition and diffraction radiation in more details. For the calculations, let's choose the parameters close to the experiment [3] ($\gamma \sim 200$ and millimeter range of radiation wave length). In Fig. 8 three curves are shown, which characterize the radial distribution of the radiation intensity on the detector situated at the distance $L$=0.5 m, for various wave lengths. With the increasing wavelength, the distribution is "smoothed" and the intensity in the maximum is suppressed because with the increasing of wavelength the parameter $R$ decreases and radius $R_{max}$ (see Fig. 6 b) increases as a result of it.

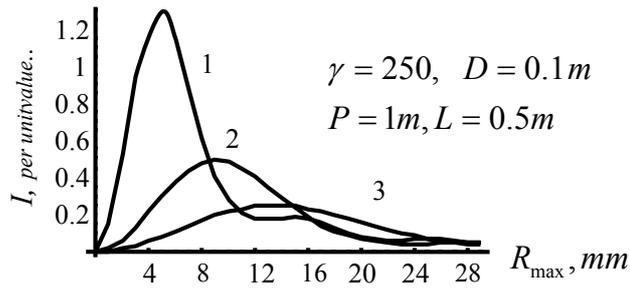

Fig. 8. The radial distribution of TR intensity from paraboloid target for various wave lengths in mm range $1 - \lambda = 1 mm; 2 - \lambda = 2 mm; 3 - \lambda = 3 mm$.

In Fig. 9, the dependence of radius of the focused spot in the point $L = P/2$ on the wave length for two target diameters is shown.

The developed approach allows the DR characteristics in paraboloid target with the hole (diameter $d_{hole}$) for the electron beam passage to be calculated. In this case the low limit in the integral (8) is $d_{hole}/2$. From Figure 10, one may see that, in our case for the $D_{hole} \leq \gamma\lambda$, the DR intensity in maximum is 20% lower than the maximum intensity of focusing TR.

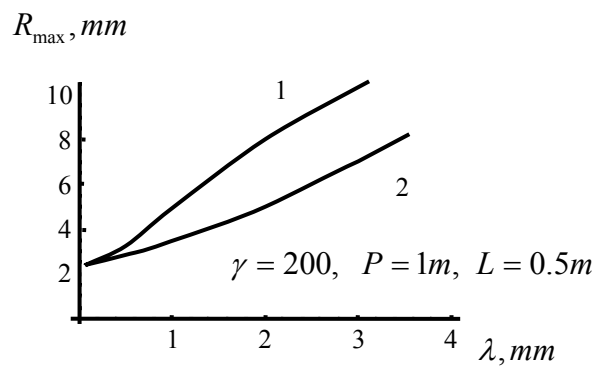

Fig. 9. The dependence on spot radius of TR from paraboloid target, on the wave length for various values of the target diameter $1 - D = 0.1 m;\ 2 - D = 0.2 m$.

In conclusion, let's note the following. The virtual photon field of ultrarelativistic particle is reflected from the deformed target surface and transformed into the real photon beam during the transition radiation process in accordance with the rules of ordinary optics that were mentioned in the work [6]. Thus, it can be expected that the use of concave targets (including the spherical ones) for the generation of TR and DR allows obtaining the focused "spot" of radiation without the use of optical system. For example, for the beam parameter in experiment [3] the paraboloid target with the focus distance of $P$=600 mm and outer diameter of 100 mm

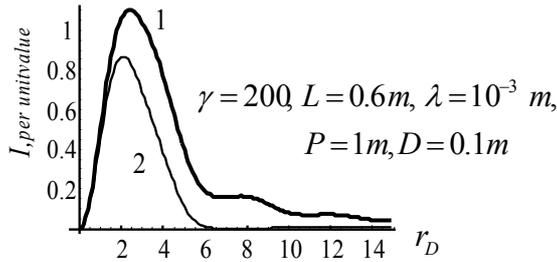

Fig. 10. The radial distribution of TR and DR intensity on the detector; 1-curve from parabolic target without the hole, 2-curve from parabolic target with the hole $D_{hole}$=10 mm

$\gamma = 200$, $L = 0.6\,m$, $\lambda = 10^{-3}\,m$,

$P = 1\,m$, $D = 0.1\,m$.

and hole one of 10 mm allows the to achieve excess of average value of the coherent diffraction radiation field $\langle E_{CTR} \rangle$ on the crystal of diameter of 10 mm situated at the distance of 300 mm which is one order higher than that one from the flat target of the same sizes. This characteristic defines the sensitivity of the technique proposed (see the expression (1)). The use of the diffraction radiation generated in the analogous target allows to avoid a deterioration of the beam parameters after its passage through the TR target (foil).

Rezaev R. was supported in part by the scholarship of the nonprofit Dynasty Foundation.